\documentclass[aps,twocolumn,showpacs]{revtex4}
\usepackage{amsmath}
\usepackage{epsfig}

\begin{document}

\title{Investigating the excited $\Omega^{0}_{c}$ states through $\Xi_{c}K$ and $\Xi^{'}_{c}K$ decay channels}

\author{Hongxia Huang$^1$, Jialun Ping$^1$\footnote{Corresponding
 author: jlping@njnu.edu.cn}, Fan Wang$^2$}

\affiliation{$^1$Department of Physics and Jiangsu Key Laboratory for Numerical
Simulation of Large Scale Complex Systems, Nanjing Normal University, Nanjing 210023, P. R. China}

\affiliation{$^2$Department of Physics, Nanjing University,
Nanjing 210093, P.R. China}

\begin{abstract}
Inspired by the five newly observed $\Omega^{0}_{c}$ states by the LHCb detector,
we study the $\Omega_{c}^{0}$ states as the $S-$wave molecular pentaquarks with
$I=0$, $J^{P}=\frac{1}{2}^{-}$, $\frac{3}{2}^{-}$, and $\frac{5}{2}^{-}$ by solving
the RGM equation in the framework of chiral quark model. Both the energies and the decay widths
are obtained in this work. Our results suggest that $\Omega_{c}(3119)^{0}$ can be explained
as an $S-$wave resonance state of $\Xi D$ with $J^{P}=\frac{1}{2}^{-}$, and the decay channels
are the $S-$wave $\Xi_{c} K$ and $\Xi^{'}_{c}K$ . Other reported $\Omega^{0}_{c}$ states cannot
be obtained in our present calculation. Another $\Omega_{c}^{0}$ state with much higher mass
3533 MeV with $J^{P}=\frac{5}{2}^{-}$ is also obtained. In addition, the calculation is extended
to the $\Omega_{b}^{0}$ states, similar results as that of $\Omega^{0}_{c}$ are obtained.
\end{abstract}


\maketitle

\setcounter{totalnumber}{5}

\section{\label{sec:introduction}Introduction}

There has been important experimental progress in the sector of heavy baryons in the past decade.
Many heavy baryons have been reported. For example, the triplet of excited $\Sigma_{c}$ baryons,
$\Sigma_{c}(2800)$, was observed by Belle~\cite{Belle1} in 2005, and they tentatively identified
the quantum numbers of these states as $J^{P}=\frac{3}{2}^{-}$. In 2008, the same neutral state
$\Sigma_{c}^{0}$ was also observed by the $BABAR$
Collaboration with the mean value of mass higher than that obtained by Belle~\cite{BABAR1}.
The charmed baryons $\Lambda_{c}(2880)^{+}$ and $\Lambda_{c}(2940)^{+}$ were observed by both
$BABAR$ and Belle Collaborations in 2007~\cite{BABAR2,Belle2}.
The charm-strange baryons $\Xi_{c}(2980)^{+}$ and $\Xi_{c}(3077)^{0}$ were reported by
Belle~\cite{Belle3} and later confirmed by $BABAR$~\cite{BABAR3}. $\Xi_{c}(3055)^{+}$ and
$\Xi_{c}(3123)^{+}$ were also investigated by $BABAR$~\cite{BABAR3}. Among the expected
charmed baryons, the spectrum of the $\Omega_{c}^{0}$ baryons, which have quark content of
$ssc$, is still unknown. Only two states: $\Omega_{c}(2695)^{0}$ and $\Omega_{c}(2768)^{0}$
with $J^{P}=1/2^{+}$ and $J^{P}=3/2^{+}$ respectively have been observed before~\cite{BABAR4,PDG2016}.
Very recently, the LHCb Collaboration reported five new narrow $\Omega_{c}^{0}$ states in the
$\Xi_{c}^{+}K^{-}$ invariant mass spectrum. They are: the $\Omega_{c}(3000)^{0}$, $\Omega_{c}(3050)^{0}$, $\Omega_{c}(3066)^{0}$, $\Omega_{c}(3090)^{0}$, and $\Omega_{c}(3119)^{0}$~\cite{LHCb}. Moreover,
the decay widths of these states were also observed by the experiment, which are only a few MeV.
However, the quantum numbers and the structures of these states are still unclear now.

All these experimental progress of heavy baryons have stimulated extensive interest in understanding the structures of the charmed baryons. A classical way to describe the charmed baryons is based on the assumption that they are conventional charmed baryons. And another way is treating them as candidates of molecular states. Take the $\Sigma_{c}(2800)$ and $\Lambda_{c}(2940)^{+}$ states for example. Being considered as two traditional charmed baryons, the strong decays of these two states have been studied by using the heavy hadron chiral perturbation theory~\cite{ChengHY}, the $^{3}P_{0}$ model~\cite{ChenC}, and the chiral quark model~\cite{ZhongXH}. On the other hand, many work treat them as candidates of molecular states. J. R. Zhang found that $\Sigma_{c}(2800)$ and $\Lambda_{c}(2940)^{+}$ as the $S-$wave $ND$ state and $ND^{*}$ state respectively by means of QCD sum rules. J. He and X. Liu explained
the $\Lambda_{c}(2940)^{+}$ as an isoscalar $S-$wave or $P-$wave $D^{*}N$ system within the one-boson-exchange model~\cite{HeJ}. For the newly observed $\Omega_{c}^{0}$ states, some work treat them as traditional charmed baryons. S. S. Agaev {\em et~al.} calculated the masses and the residues of these states with $J^{P}=1/2^{+}$ and $J^{P}=3/2^{+}$ in the framework of QCD two-point sum rules and they were inclined to assign the $\Omega_{c}(2066)^{0}$ and $\Omega_{c}(3119)^{0}$ states as the first radially excited ($2S,1/2^{+}$) and ($2S,3/2^{+}$) charmed baryons~\cite{Agaev}. H. X. Chen {\em et~al.} studied the decay properties of the $P-$wave charmed baryons within the light-cone QCD sum rules, including some $\Lambda_{c}$, $\Sigma_{c}$, $\Xi_{c}$ states, as well as these newly reported $\Omega_{c}^{0}$ states~\cite{HXChen}. They interpreted one of these $\Omega_{c}$ states was a $J^{P}=1/2^{-}$ state, two of them were $J^{P}=3/2^{-}$ state and $J^{P}=5/2^{-}$ state, another two may be with $J^{P}=1/2^{+}$ and $3/2^{+}$. M. Karliner and J. L. Rosner explained these $\Omega_{c}^{0}$ baryons as bound states of a $P-$wave $ss-$diquark and a $c-$quark~\cite{Karliner}, and they predicted two of spin $1/2$, two of spin $3/2$, and one of spin $5/2$, all with negative parity. K. L. Wang {\em et~al.} investigated the strong and radiative decay properties of the low-lying $\Omega_{c}$ states in a constituent quark model~\cite{KLWang}. Their results show that the $\Omega_{c}(3000)^{0}$ and $\Omega_{c}(3090)^{0}$ can be assigned to have $J^{P}=1/2^{-}$, the $\Omega_{c}(3050)^{0}$ with $J^{P}=3/2^{-}$, the $\Omega_{c}(3066)^{0}$ with $J^{P}=5/2^{-}$, and the $\Omega_{c}(3119)^{0}$ might be one of the two $2S$ states of the first radial excitations.  Another way to describe these states is assuming they are pentaquark states. G. Yang {\em et~al.} did a dynamical calculation of 5-quark systems to study the structure of the pentaquarks $\Omega_{c}$ in the chiral quark model by taking the advantage of gaussian expansion method~\cite{GYang}, and they pointed out that the $\Xi\bar{D}$, $\Xi_{c}\bar{K}$ and $\Xi_{c}^{*}\bar{K}$ are possible the candidates of these new particles.

Actually, the hadron-hadron scattering is one of the important ways to generate and identify multi-quark states. Therefore, to provide the necessary information for experiment to search for the multi-quark states, we should not only calculate the mass spectrum but also study the corresponding scattering process. The scattering phase shifts will show a resonance behavior in the resonance energy region. By using the constituent quark models and the resonating group method (RGM)~\cite{RGM}, we have obtained the $d^{*}$ resonance during the $NN$ scattering process, the energy and decay width of the partial wave are consistent with the experiment data~\cite{Ping2009}. Extending to the pentaquark system, we investigated the $N\phi$ state in the different scattering channels: $N\eta'$, $\Lambda K$, and $\Sigma K$~\cite{Gao2017}. Both the resonance mass and decay width were obtained, which provide the necessary information for experiment searching at Jefferson Lab. Therefore, it is interesting to extend such study to the newly observed $\Omega_{c}^{0}$ states. In this work, we will assume $\Omega_{c}^{0}$ states are pentaquark states, calculate both the masses and decay widths of these states, and analyze if there are some $\Omega_{c}^{0}$ states which can be explained as pentaquarks by comparing with the LHCb data. Finally, we will also extend the study to the $\Omega_{b}^{0}$ states because of the heavy flavor symmetry.

The structure of this paper is as follows. A brief introduction of
a constituent quark model used is given in section II. Section III devotes to the numerical
results and discussions. The summary is shown in the last section.

\section{Chiral quark model}

Here, we use the chiral quark model to study the $\Omega_{c}^{0}$ states. The Salamanca model was chosen as the representative of the chiral quark models, because the Salamanca group's work covers the hadron spectra, nucleon-nucleon interaction, and
multiquark states. We also have used this model to study the nucleon-nucleon interaction, dibaryon resonance states, such as $d^{*}$ state~\cite{Ping2009}, $N\Omega$~\cite{ChenM2011,Huang2015}, and so on. In this model,the constituent quarks interact with each other through the one-gluon-exchange and
the Goldstone boson exchange in addition to the color confinement. For the system with strangeness, a version of chiral quark model~\cite{Garcilazo,QBLi}
had been used, where full SU(3) scalar octet meson-exchange was used.
These scalar potentials have the same functional form as the one of
SU(2) ChQM but a different SU(3) operator dependence~\cite{Garcilazo}.
The model details can be found in Ref.~\cite{Salamanca}. Here we only give the Hamiltonian:
\begin{widetext}
\begin{eqnarray}
H & = & \sum_{i=1}^6\left(m_i+\frac{p_i^2}{2m_i}\right)-T_{CM} +\sum_{j>i=1}^6
\left(V^{C}_{ij}+V^{G}_{ij}+V^{\chi}_{ij}+V^{\sigma}_{ij}\right), \\
V^{C}_{ij} & = & -a_{c} \boldsymbol{\lambda}^c_{i}\cdot \boldsymbol{
\lambda}^c_{j} ({r^2_{ij}}+v_{0}), \label{sala-vc} \\
V^{G}_{ij} & = & \frac{1}{4}\alpha_s \boldsymbol{\lambda}^{c}_i \cdot
\boldsymbol{\lambda}^{c}_j
\left[\frac{1}{r_{ij}}-\frac{\pi}{2}\delta(\boldsymbol{r}_{ij})(\frac{1}{m^2_i}+\frac{1}{m^2_j}
+\frac{4\boldsymbol{\sigma}_i\cdot\boldsymbol{\sigma}_j}{3m_im_j})-\frac{3}{4m_im_jr^3_{ij}}
S_{ij}\right] \label{sala-vG} \\
V^{\chi}_{ij} & = & V_{\pi}( \boldsymbol{r}_{ij})\sum_{a=1}^3\lambda
_{i}^{a}\cdot \lambda
_{j}^{a}+V_{K}(\boldsymbol{r}_{ij})\sum_{a=4}^7\lambda
_{i}^{a}\cdot \lambda _{j}^{a}
+V_{\eta}(\boldsymbol{r}_{ij})\left[\left(\lambda _{i}^{8}\cdot
\lambda _{j}^{8}\right)\cos\theta_P-(\lambda _{i}^{0}\cdot
\lambda_{j}^{0}) \sin\theta_P\right] \label{sala-Vchi1} \\
V_{\chi}(\boldsymbol{r}_{ij}) & = & {\frac{g_{ch}^{2}}{{4\pi
}}}{\frac{m_{\chi}^{2}}{{\
12m_{i}m_{j}}}}{\frac{\Lambda _{\chi}^{2}}{{\Lambda _{\chi}^{2}-m_{\chi}^{2}}}}%
m_{\chi} \left\{(\boldsymbol{\sigma}_{i}\cdot
\boldsymbol{\sigma}_{j})
\left[ Y(m_{\chi}\,r_{ij})-{\frac{\Lambda_{\chi}^{3}}{m_{\chi}^{3}}}%
Y(\Lambda _{\chi}\,r_{ij})\right] \right.\nonumber \\
&& \left. +\left[H(m_{\chi}
r_{ij})-\frac{\Lambda_{\chi}^3}{m_{\chi}^3}
H(\Lambda_{\chi} r_{ij})\right] S_{ij} \right\}, ~~~~~~\chi=\pi, K, \eta, \\
V^{\sigma_{a}}_{ij} & = & V_{a_{0}}(
\boldsymbol{r}_{ij})\sum_{a=1}^3\lambda _{i}^{a}\cdot \lambda
_{j}^{a}+V_{\kappa}(\boldsymbol{r}_{ij})\sum_{a=4}^7\lambda
_{i}^{a}\cdot \lambda _{j}^{a}+V_{f_{0}}(\boldsymbol{r}_{ij})\lambda _{i}^{8}\cdot \lambda
_{j}^{8}+V_{\sigma}(\boldsymbol{r}_{ij})\lambda _{i}^{0}\cdot
\lambda _{j}^{0} \label{sala-su3} \\
V_{k}(\boldsymbol{r}_{ij}) & = & -{\frac{g_{ch}^{2}}{{4\pi }}}
{\frac{\Lambda _{k}^{2}m_{k}}{{\Lambda_{k}^{2}-m_{k}^{2}}}}%
\left[ Y(m_{k}\,r_{ij})-{\frac{\Lambda _{k}}{m_{k}}}%
Y(\Lambda _{k}\,r_{ij})\right] , ~~~~~~k=a_{0}, \kappa, f_{0}, \sigma,  \\
S_{ij}&=&\left\{ 3\frac{(\boldsymbol{\sigma}_i
\cdot\boldsymbol{r}_{ij}) (\boldsymbol{\sigma}_j\cdot
\boldsymbol{r}_{ij})}{r_{ij}^2}-\boldsymbol{\sigma}_i \cdot
\boldsymbol{\sigma}_j\right\},\\
H(x)&=&(1+3/x+3/x^{2})Y(x),~~~~~~
 Y(x) =e^{-x}/x. \label{sala-vchi2}
\end{eqnarray}
\end{widetext}
Where $S_{ij}$ is quark tensor operator; $Y(x)$ and $H(x)$ are
standard Yukawa functions; $T_c$ is the kinetic
energy of the center of mass; $\alpha_{s}$ is the quark-gluon coupling constant; $g_{ch}$ is the coupling constant
for chiral field, which is determined from the $NN\pi$ coupling constant through
\begin{equation}
\frac{g_{ch}^{2}}{4\pi }=\left( \frac{3}{5}\right) ^{2}{\frac{g_{\pi NN}^{2}%
}{{4\pi }}}{\frac{m_{u,d}^{2}}{m_{N}^{2}}}\label{gch}.
\end{equation}
The other symbols in the above expressions have their usual meanings.

Generally, we use the parameters from our former work of dibaryons, only the mass of charm quark is adjusted to fit the charmed mesons and baryons used in this work. However, the former parameters can describe the ground baryons well, but cannot fit the ground mesons, especially the $K$ meson, the mass of which is much higher than the experimental value. This situation will lead to a consequence that some bound states cannot decay to the open channel $\Xi_{c} K$, because of the much larger mass of $K$. To solve this problem, we adjust the parameters which are related to $s$ and $c$ quarks in this work, and keep the parameters which are related to $u$ and $d$ quarks. By doing this, the parameters can describe the nucleon-nucleon interaction well, and at the same time, it will lower the mass of $K$. All the parameters of Hamiltonian are given in Table~\ref{parameters}. The calculated masses of baryons and mesons in comparison with experimental values are shown in Table~\ref{mass}.

\begin{table}[ht]
\caption{\label{parameters}Model parameters:
$m_{\pi}=0.7$fm$^{-1}$, $m_{ k}=2.51$fm$^{-1}$,
$m_{\eta}=2.77$fm$^{-1}$, $m_{\sigma}=3.42$fm$^{-1}$,
$m_{a_{0}}=m_{\kappa}=m_{f_{0}}=4.97$fm$^{-1}$,
$\Lambda_{\pi}=4.2$fm$^{-1}$, $\Lambda_{K}=5.2$fm$^{-1}$,
$\Lambda_{\eta}=5.2$fm$^{-1}$, $\Lambda_{\sigma}=4.2$fm$^{-1}$,
$\Lambda_{a_{0}}=\Lambda_{\kappa}=\Lambda_{f_{0}}=5.2$fm$^{-1}$,
$g_{ch}^2/(4\pi)$=0.54, $\theta_p$=$-15^{0}$. }
\begin{tabular}{ccccccccc} \hline\hline
$b$ & ~~~$m_{u}$~~~~ & ~~~$m_{d}$~~~ & ~~~$m_{s}$~~~ & ~~~$m_{c}$~~~ & ~~~$m_{b}$~~~~   \\
(fm) & (MeV) & (MeV) & (MeV) & (MeV) & (MeV)   \\ \hline
0.518  & 313 & 313 &  450 & 1635 &    4988  \\ \hline
 $ a_c$ & $V_{0}$ & $\alpha_{s_{uu}}$ &  $\alpha_{s_{us}}$ & $\alpha_{s_{uc}}$ & $\alpha_{s_{ub}}$  \\
(MeV\,fm$^{-2}$) & (MeV) & & & & & \\ \hline
  48.59 & -0.961 & 0.84 & 0.82 & 0.35 & 0.27 &\\ \hline
 $\alpha_{s_{ss}}$ & $\alpha_{s_{sc}}$ & $\alpha_{s_{sb}}$ & $\alpha_{s_{cc}}$ & $\alpha_{s_{bb}}$ & \\\hline
 0.30 & 0.25 & 0.20 & 0.20 & 0.15 & \\
\hline\hline
\end{tabular}
\end{table}

\begin{table}[ht]
\caption{\label{mass}The masses of baryons and mesons used in this work (in MeV).}
\begin{tabular}{lcccccccc}
\hline \hline
           &  ~~$\Xi$~~  & ~~$\Xi^*$~~  &  ~~$\Xi_{c}$~~ & ~~$\Xi_{c}^{'}$~~  & ~~$\Xi_{c}^*$~~  & ~~$\Omega_{c}$~~   & ~~$\Omega_{c}^{*}$~~  \\ \hline
Exp.       &1318 &1533 &2469 &2577 &2646 &2695 &2766  \\
ChQM      &1225 &1359 &2448 &2527 &2543 &2662 &2672  \\  \hline
          &  ~~$K$~~  & ~~$K^*$~~  &  ~~$\eta$~~ & ~~$\omega$~~  & ~~$D$~~  & ~~$D^{*}$~~   & ~~~~  \\
Exp.       &495 &892 &548 &783 &1864 &2007 &  \\
ChQM      &593 &830 &523 &702 &1980 &2007 &  \\  \hline
            & ~~$\Xi_{b}$~~ & ~~$\Xi_{b}^{'}$~~  & ~~$\Xi_{b}^*$~~  & ~~$\Omega_{b}$~~   & ~~$\Omega_{b}^{*}$~~ & ~~$B$~~  & ~~$B^{*}$~~ \\ \hline
Exp.       &5795 &5935 &5949 &6046 &? &5279 &5325  \\
ChQM      &5794 &5880 &5884 &6008 &6011 &5351 &5358  \\  \hline
 \hline
\end{tabular}
\end{table}

\section{The results and discussions}

In this work, we investigate the $S-$wave $\Omega_{c}^{0}$ states as the molecular pentaquarks with $I=0$, 
$J^{P}=\frac{1}{2}^{-}$, $\frac{3}{2}^{-}$, and $\frac{5}{2}^{-}$. Three structures are considered here, 
and they are structure 1: $uss-c\bar{u}$, 2: $usc-s\bar{u}$, and 3: $ssc-u\bar{u}$. All the channels 
involved are listed in Table~\ref{channels}.

\begin{table}
\caption{The channels calculated in this work.}
\begin{tabular}{lcccccccccc}
\hline \hline
~~$J^{P}$~~  & ~~~~~~Structure~~~~~~ &  ~~~~~~~~~Channels~~~~~~~~~~ \\  \hline
 ~~$\frac{1}{2}^{-}$ & 1. $uss-c\bar{u}$   & $\Xi D$, $\Xi D^{*}$, $\Xi^{*}D^{*}$    \\
                   & 2. $usc-s\bar{u}$  & $\Xi_{c}K$, $\Xi^{'}_{c}K$, $\Xi_{c}K^{*}$, $\Xi^{'}_{c}K^{*}$, $\Xi^{*}_{c}K^{*}$    \\
                   & 3. $ssc-u\bar{u}$   & $\Omega_{c}\eta$, $\Omega_{c}\omega$, $\Omega^{*}_{c}\omega$    \\  \hline
 ~~$\frac{3}{2}^{-}$ & 1. $uss-c\bar{u}$   & $\Xi D^{*}$, $\Xi^{*} D$, $\Xi^{*}D^{*}$    \\
                   & 2. $usc-s\bar{u}$  & $\Xi_{c}K^{*}$, $\Xi^{'}_{c}K^{*}$, $\Xi^{*}_{c}K$, $\Xi^{*}_{c}K^{*}$    \\
                   & 3. $ssc-u\bar{u}$   & $\Omega_{c}\omega$, $\Omega^{*}_{c}\eta$, $\Omega^{*}_{c}\omega$    \\ \hline
 ~~$\frac{5}{2}^{-}$ & 1. $uss-c\bar{u}$   & $\Xi^{*}D^{*}$    \\
                   & 2. $usc-s\bar{u}$  & $\Xi^{*}_{c}K^{*}$    \\
                   & 3. $ssc-u\bar{u}$   & $\Omega^{*}_{c}\omega$    \\
 \hline\hline
\end{tabular}
\label{channels}
\end{table}

\subsection{Bound state calculation}

As the first step, we do a dynamic calculation based on RGM to check whether or not there is any bound state. 
We expand the relative motion wavefunction between two clusters in the RGM equation by Gaussian bases. 
By doing this, the integro-differential equation of RGM can be reduced to algebraic equation, generalized
eigen-equation. Then we can obtain the energy of the system by solving this generalized eigen-equation. 
In the present calculation, the baryon-meson separation is taken to be less than 6 fm (to keep the dimensions 
of matrix manageably small). The single channel calculation shows that the energy of each channel locates above 
the threshold of the corresponding channel, which means that there is no any singlet bound state. 
By coupling all the channels with different structures, there exist some bound states. The binding energies and 
the masses of the bound states, as well as the percentages of each channel in the eigen-states are listed in 
Table~\ref{bound}. Before discussing the features of the states, we should mention how we obtain the mass of 
these states. The binding energy $B=M^{the.}-M^{the.}_{B}-M^{the.}_{M}$, where $M^{the.}$, $M^{the.}_{B}$ and 
$M^{the.}_{M}$ stand for the theoretical mass of the molecular state, a baryon and a meson, respectively. 
To minimize the theoretical errors and to compare calculated results to the experimental data, we shift the mass 
of a molecular state to $M=M^{exp.}_{B}+M^{exp.}_{M}+B$, where the experimental values of a baryon and a meson 
are used. Taking the state $J^P={\frac12}^{-}~\Xi D$ an example, the calculated mass of this state is $3169$ MeV, 
then the binding energy $B$ is obtained by subtracting the theoretical masses of $\Xi$ and $D$, 
$3169-1225-1980=-36$ (MeV). Adding the experimental masses of the hadrons, the mass of this state 
$M=1318+1864+(-36)=3146$ (MeV) is arrived.

\begin{table}
\begin{center}
\caption{The binding energy and masses (in MeV) of the molecular pentaquarks with channel-coupling
 and the percentages of each channel in the eigen-states.}
{\begin{tabular}{@{}cc|cc|cc@{}c} \hline
\multicolumn{2}{c|}{$J^{P}=\frac{1}{2}^{-}$}
 &\multicolumn{2}{c|}{$J^{P}=\frac{3}{2}^{-}$}&\multicolumn{2}{c}{$J^{P}=\frac{5}{2}^{-}$}\\ \hline
 ~~$E_B$~~ & $-36$~~ & ~~$E_B$~~ & $-1$~~ & ~~$E_B$~~ & $-7$~~   \\ 
 ~~$M_{cc}$~~ & ~~~~3146~~~~ & ~~$M_{cc}$~~ & ~~~~3324~~~~ & ~~$M_{cc}$~~ & ~~~~3533~~~~   \\ \hline
 {$\Xi D$}      & 58.1 &  {$\Xi D^{*}$} & 63.6   & {$\Xi^{*}D^{*}$} & 50.3 \\
 {$\Xi D^{*}$}  & 3.3  & {$\Xi^{*}D$} & 3.6   & {~$\Xi_{c}^{*}K^{*}$~} & 13.4 \\
 {~$\Xi^{*} D^{*}$~}  & 1.9  & {~$\Xi^{*}D^{*}$~} &  0.2   & {$\Omega^{*}_{c}\omega$} & 36.3  \\
 {$\Omega_{c}\eta$}  & 31.1  & {$\Omega_{c}\omega$} &  1.2 & &    \\
 {$\Omega_{c}\omega$}   &  1.9  & {$\Omega^{*}_{c}\eta$} &  31.3 & &   \\
 {$\Omega^{*}_{c}\omega$}   & 3.7  & {$\Omega^{*}_{c}\omega$} & 0.1 &   &      \\
 \hline
\end{tabular}
\label{bound}}
\end{center}
\end{table}

For the $J^{P}=\frac{1}{2}^{-}$ system, the channel-coupling of each structure cannot make any state bound. 
Then we do a channel-coupling among different structures. We find that there is no bound state by coupling channels 
of structure 1 and 2, or structures 2 and 3. However, by coupling the channels of structure 1 and 3, we obtain a 
stable state, the mass of which is $3146$ MeV, $-36$ MeV lower than the threshold of $\Xi D$, and the main component 
of this state is $\Xi D$, as showed in Table ~\ref{bound}. While coupling these channels to the channels of structure 2, 
in which there are open channels $\Xi_{c} K$ and $\Xi^{'}_{c}K$, whose thresholds are lower than the mass of $\Xi D$. 
The result of the channel-coupling of three structures shows that the lowest eigen-energy is very close but still higher 
than the threshold of the $\Xi_{c} K$, which means there is no bound state by all channels coupling. However, we also 
obtain a quasi-stable state, the mass of which is smaller than the threshold of $\Xi D$, but it fluctuates around the 
$3146$ MeV with about 1 MeV with the variation of the baryon-meson separation. To confirm whether or not the state 
$\Xi D$ can survive as a resonance state after the full channels coupling, the study of the scattering process of the 
open channels is needed, which is discussed in subsection B.

For the $J^{P}=\frac{3}{2}^{-}$ system, we only find a bound state $\Xi D^{*}$ with a binding energy of only $-1$ MeV 
by coupling channels of structure 1 and 3, as shown in Table ~\ref{bound}. There is neither any bound state nor any 
quasi-stable state by the full channels coupling. However, we still need to calculate the scattering process of the open 
channel $\Xi^{*}_{c}K$ to check if the state $\Xi D^{*}$ is a resonance or not. The result is also shown in subsection B.

For the $J^{P}=\frac{5}{2}^{-}$ system, it includes only one channel of each structure. Although it is not bound for each 
structure, there exists a bound state by three channels coupling. The binding energy and the mass of this system, as well as 
the percentages of each channel in the eigen-state are shown in Table ~\ref{bound}, from which we can see that the mass of 
the $J^{P}=\frac{5}{2}^{-}$ system is $3533$ MeV. Moreover, this state can also decay to some open channels, but they are 
$D-$wave channels. This $S-$ and $D-$wave channel-coupling, which is through the tenser force, is always very weak in our 
quark model calculation~\cite{Gao2017}. So we can estimate the effect of this kind of coupling is small here. We will do 
this $S-$ and $D-$wave channel-coupling in future.

\subsection{Resonance states and decay widths}

To find the resonance mass and decay width of the quasi-stable states discussed in subsection A, we calculate the phase shifts 
of the corresponding open channels. For the $J^{P}=\frac{1}{2}^{-}$ system, coupling to the $S-$wave open channel $\Xi_{c} K$ 
cause the $\Xi D$ bound state to change into an elastic resonance, where the phase shift, shown in Fig. 1, rises through 
$\pi$ at a resonance mass. We find the resonance mass is $3146.8$ MeV, which shows the energy of the bound state is pushed up 
a little. From the Fig. 1, the decay width is obviously very narrow, which is only $0.6$ MeV. By coupling to another open 
channel $\Xi^{'}_{c} K$, similar results are obtained. The $\Xi D$ bound state changes to a resonance state of the same 
resonance mass and decay width. Therefore, we can obtain a resonance state $\Xi D$ with $J^{P}=\frac{1}{2}^{-}$ in the 
decay channel $\Xi_{c} K$ or $\Xi^{'}_{c} K$, with the resonance mass $3146.8$ MeV and decay width $0.6$ MeV, which is 
consistent with the newly reported $\Omega_{c}(3119)^{0}$, the decay width of which is $1.1\pm0.8\pm0.4$ MeV. What's more, 
this $\Omega_{c}(3119)^{0}$ were observed both in $\Xi_{c} K$ and $\Xi^{'}_{c} K$ in the LHCb experiment~\cite{LHCb}. 
So in our quark model calculation, we can explain the $\Omega_{c}(3119)^{0}$ as a resonance state $\Xi D$ with 
$J^{P}=\frac{1}{2}^{-}$.

\begin{figure}[ht]
\begin{center}
\epsfxsize=3.0in \epsfbox{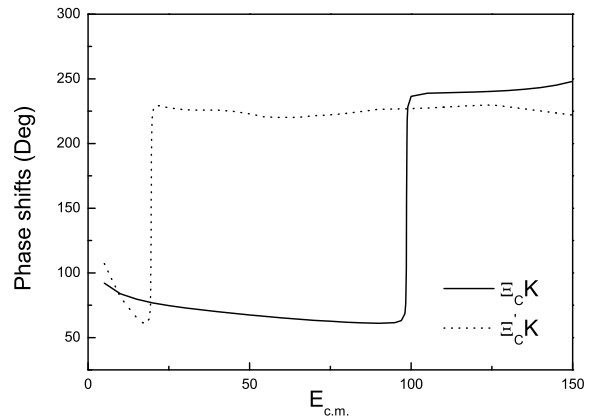} \vspace{-0.1in}

\caption{The phase shifts of the scattering channels $\Xi_{c} K$ and $\Xi^{'}_{c} K$ for the $J^{P}=\frac{1}{2}^{-}$ system.}
\end{center}
\end{figure}

\begin{figure}[ht]
\begin{center}
\epsfxsize=3.0in \epsfbox{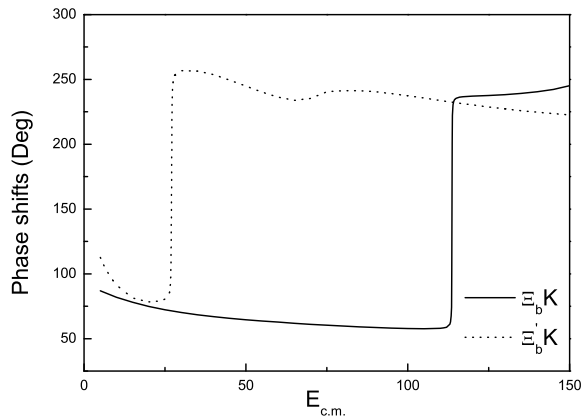} \vspace{-0.1in}

\caption{The phase shifts of the scattering channels $\Xi_{b} K$ and $\Xi^{'}_{b} K$ for the $J^{P}=\frac{1}{2}^{-}$ system.}
\end{center}
\end{figure}

For the $J^{P}=\frac{3}{2}^{-}$ system, we study the scattering process of the open channel $S-$wave $\Xi^{*}_{c}K$, and 
we do not find any resonance state. This is reasonable. Because the binding energy of the $\Xi D^{*}$ is only $-1$ MeV, 
which is too small. The channel-coupling to the $\Xi^{*}_{c}K$ pushes the energy of $\Xi D^{*}$ above its threshold. 
So there is no resonance state with $J^{P}=\frac{3}{2}^{-}$ in our calculation.

In addition, we also extend the study to the $\Omega^{0}_{b}$ system because of the heavy flavor symmetry. The results 
are similar to that of $\Omega^{0}_{c}$ system. We obtain a resonance state $\Xi B$ with $J^{P}=\frac{1}{2}^{-}$ in the 
decay channels $\Xi_{b} K$ and $\Xi^{'}_{b} K$, with the same resonance mass of $6560$ MeV and decay width $0.6$ MeV 
(see Fig. 2). Besides, there is a bound state $\Xi^{*}B^{*}$ with $J^{P}=\frac{5}{2}^{-}$ and the energy of it is $6856$ MeV.

\section{Summary}
In summary, we investigate the excited $\Omega_{c}^{0}$ states as the $S-$wave molecular pentaquarks with $I=0$, 
$J^{P}=\frac{1}{2}^{-}$, $\frac{3}{2}^{-}$, and $\frac{5}{2}^{-}$ by solving the RGM equation in the framework of 
chiral quark model. In this work, we study not only the energies of the $\Omega_{c}^{0}$ states but also the decay 
widths of them. Our results show that the $\Omega_{c}(3119)^{0}$ can be explained as an $S-$wave resonance state 
$\Xi D$ with $J^{P}=\frac{1}{2}^{-}$, and the decay channels are the $S-$wave $\Xi_{c} K$ and $\Xi^{'}_{c} K$. 
Other newly reported $\Omega^{0}_{c}$ states cannot be obtained in our present calculation. They maybe the conventional 
charmed baryons with $P-$wave or even higher partial waves. It is also possible that these states are mixing states 
with $q^{3}$ and $q^{4}\bar{q}$. The unquenched quark model, which takes into account the high Fock components, 
is feasible to do this mixing. We also obtain another $\Omega_{c}^{0}$ state with much higher mass, which is $3533$ MeV 
with $J^{P}=\frac{5}{2}^{-}$. Besides, the calculation is extended to the $\Omega_{b}^{0}$ states, similar results as 
that of $\Omega^{0}_{c}$ are obtained.

Our calculation also shows that the coupling of the structure $uss-c\bar{u}$ and $ssc-u\bar{u}$ is important to make 
the $\Xi D$ bound, and the coupling with structure $usc-s\bar{u}$ is very weak, which can be used to explain why the 
reported $\Omega_{c}(3119)^{0}$ has a narrow decay width. It also gives us some information that the decay width of 
these $\Omega_{c}^{0}$ states maybe somehow related to the structure of these states. More structures will be studied 
in future.

\acknowledgments{This work is supported partly by the National Science Foundation
of China under Contract Nos. 11675080, 11035006 and 11535005, the Natural Science Foundation of
the Jiangsu Higher Education Institutions of China (Grant No. 16KJB140006), and Jiangsu Government
Scholarship for Overseas Studies.}


\begin{thebibliography}{99}
\bibitem{Belle1} R. Mizuk {\em et al.} (Belle Collaboration), Phys. Rev. Lett. {\bf 94}, 122002 (2005).
\bibitem{BABAR1} B. Aubert {\em et al.} (BABAR Collaboration), Phys. Rev. D {\bf 78}, 112003 (2008).
\bibitem{BABAR2} B. Aubert {\em et al.} (BABAR Collaboration), Phys. Rev. Lett. {\bf 98}, 012001 (2007).
\bibitem{Belle2} R. Mizuk {\em et al.} (Belle Collaboration), Phys. Rev. Lett. {\bf 98}, 262001 (2007).
\bibitem{Belle3} R. Chistov {\em et al.} (Belle Collaboration), Phys. Rev. Lett. {\bf 97}, 162001 (2006).
\bibitem{BABAR3} B. Aubert {\em et al.} (BABAR Collaboration), Phys. Rev. D {\bf 77}, 012002 (2008).
\bibitem{BABAR4} B. Aubert {\em et al.} (BABAR Collaboration), Phys. Rev. Lett. {\bf 97}, 232001 (2006).
\bibitem{PDG2016} Particle Data Group, C. Patrignani {\em et al.}, Chin. Phys. C {\bf 40}, 100001 (2016).
\bibitem{LHCb} R. Aaij {\em et al.} (LHCb Collaboration), arXiv:1703.04639v1 [hep-ex].
\bibitem{ChengHY} H. Y. Cheng and C. K. Chua, Phys. Rev. D {\bf 75}, 014006 (2007).
\bibitem{ChenC} C.Chen, X. L.Chen, X. Liu, W. Z. Deng and S. L. Zhu, Phys. Rev. D {\bf 75}, 094017 (2007).
\bibitem{ZhongXH} X. H. Zhong and Q. Zhao, Phys. Rev. D {\bf 77}, 074008 (2008).
\bibitem{HeJ} J. He and X. Liu, Phys. Rev. D {\bf 82}, 114029 (2010).
\bibitem{Agaev} S. S. Agaev, K. Azizi and H. Sundu, arXiv:1703.07091v1 [hep-ph].
\bibitem{HXChen} H. X. Chen, Q. Mao, W. Chen, A. Hosaka, X. Liu and S. L. Zhu, arXiv:1703.07703v1 [hep-ph].
\bibitem{Karliner} M. Karliner and J. L. Rosner, arXiv:1703.07774v1 [hep-ph].
\bibitem{KLWang} K. L. Wang, L. Y. Xiao, X. H. Zhong and Q. Zhao, arXiv:1703.09130v1 [hep-ph].
\bibitem{GYang} G. Yang and J. L. Ping, arXiv:1703.08845v1 [hep-ph].
\bibitem{RGM} M. Kamimura, Supp. Prog. Theo. Phys. {\bf 62}, 236 (1977).
\bibitem{Ping2009} J. L. Ping, H. X. Huang, H. R. Pang, F. Wang and C. W. Wong, Phys. Rev.
C {\bf 79}, 024001 (2009).
\bibitem{Gao2017} H. Gao, H. X. Huang, T. B. Liu, J. L. Ping, F. Wang and Z. Zhao, arXiv:1701.03210v1 [hep-ph].
\bibitem{ChenM2011} M. Chen, H. X. Huang, J. L. Ping and F. Wang, Phys. Rev. C {\bf 83}, 015202 (2011).
\bibitem{Huang2015} H. X. Huang, J. L. Ping and F. Wang, Phys. Rev. C {\bf 92}, 065202 (2015).
\bibitem{Garcilazo} H. Garcilazo, T. F. Carames and A. Valcarce, Phys. Rev. C {\bf 74}, 034002 (2007).
\bibitem{QBLi}Q. B. Li, P. N. Shen, Z. Y. Zhang and Y. W. Yu, Nucl. Phys. A {\bf 683}, 487 (2001).
\bibitem{Salamanca} A. Valcarce, H. Garcilazo, F. Fern\'{a}ndez and P. Gonzalez,
  Rep. Prog. Phys. {\bf 68}, 965 (2005) and references therein.
\end{thebibliography}
\end{document}